\documentclass[showpacs,showkeys,preprintnumbers,amsmath,aps]{revtex4}
\usepackage{epsfig}
\begin{document}

\markboth{Zhaoxin Xu} {QMC simulations on AFM-FM random
alternating spin chain}

%\catchline{}{}{}{}{}
%
%%%%%%%%%%%%%%%%%%%%%%%%%%%%%%%%%%%%%%%%%%%%%%%%%%%%%%%%%%%%%%%%%%%%

\title{Quantum Monte Carlo Simulations on S=1/2 \\
Antiferromagnetic-Ferromagnetic Random Alternating spin chain}

\author{\footnotesize Peng Zhang,\quad Zhaoxin
  Xu\footnote{E-mail:zxxu@zimp.zju.edu.cn},\quad Heping Ying and \quad Jianhui Dai}

\affiliation{Zhejiang Institute of Modern Physics, Zhejiang University,\\
Hangzhou, 310027,
P.R. China}

%\maketitle

%\pub{\today}
\date{\today}

\begin{abstract}
The S=1/2 Heisenberg chain with bond alternation and randomness of
antiferromagnetic (AFM) and ferromagnetic (FM) interactions is
investigated by quantum Monte Carlo simulations of loop/cluster
algorithm. Our results have shown interesting finite temperature magnetic
properties of this model. The relevance of our study to the observed
results of the material (CH$_3$)$_2$CHNH$_3$Cu(Cl$_x$Br$_{1-x}$)$_3$
is discussed.

\end{abstract}

\maketitle

\section{Introduction}
Randomness induced quantum phase transitions have attracted
intensive interests in the past decades. Putting enough strong bond
randomness, analyses of real space renormalization group (RSRG)
method have shown that RG flows of spin S=1 and S=1/2 quantum
antiferromagnetic Heisenberg chains \cite{Dasgupta,Fisher,Hyman1}
go to a stable fixed point called the random singlet (RS) phase in
which spins far apart in space randomly form weakly bound singlet
pairs. This property induces universal behaviors of ground
states and low temperature thermodynamics, e.g. the
energy spectrum is gapless, the temporal correlation length
$\xi_{\tau}$ and spatial correlation length $\xi_L$ diverge at
zero temperature, and there is a non-universal infinite dynamical
exponent $z$ which comes from the relation $\xi_{\tau}^z \sim
\xi_L$. More important, the uniform
susceptibility diverges universally in the RS phase as
\begin{equation}
\chi \sim \frac{1}{T {\bf ln}^2 (\Omega/T)}
\label{eq_RSxu}
\end{equation}
at low temperature, where $\Omega$ is a non-universal constant.
On the other hand, the S=1/2 dimerized AFM chain was found to be extremely stable
against strong bond randomness\cite{Hyman2}. In this case, the system is in a
quantum Griffiths-McCoy (QG) phase when the bond randomness is strong
enough. This phase is characterized by gapless excitations and
finite correlation length. In the QG phase, the uniform
susceptibility behaves as
\begin{equation}
\chi \sim T^{-\gamma}
\label{eq_QGxu}
\end{equation}
at low temperature, where $\gamma$ is a non-universal exponent.
On the experiments, the bond randomness effects have been found in several
antiferromagnetic quasi-1D materials
%such as
%BaCu$_2$(Si$_{x}$Ge$_{1-x}$)$_2$O$_7$, Mg-doped
%PbNi$_2$V$_2$O$_8$, and quasi-1D S=1/2 spin ladder material
%Zn-doped SrCu$_2$O$_3$
 \cite{Uchiyama,Azuma,Masuda}. Especially, a 
recent experiment on BaCu$_2$(Si$_{0.5}$Ge$_{0.5}$)$_2$O$_7$
\cite{Masuda} clearly show typical scaling relations of RS phase
predicted by theory\cite{Damle}.
Moreover, there is another kind of bond randomness whose bonds can be both
AFM and FM. In such systems, the RSRG analyses
\cite{Westerberg,Frischmuth} predicted a
universal fixed point different from the RS phase because the
spins correlate to form effective spins whose average size grows
with lowering of the energy scale, 
%Candidate materials of such
%randomness, e.g. Sr$_3$CuPt$_{1-x}$Ir$_x$O$_6$ \cite{Nguyen}, is
%considered to contain both AFM and FM bonds whose fraction is
%related to the concentration $x$ of Ir.
 and the magnetic susceptibility is {\it Curie-like}
 $\chi_u \sim 1/T$. Materials with
 such randomness is also fabricated \cite{Nguyen}.

Besides the above mentioned two kinds of bond randomness, Manaka and
coworkers recently found that the compound
(CH$_3$)$_2$CHNH$_3$Cu(Cl$_x$Br$_{1-x}$)$_3$ \cite{Manaka1} can be
considered as a bond randomness S=1/2 AFM and FM
alternating Heisenberg chains. The isomorphous compounds
(CH$_3$)$_2$CHNH$_3$CuCl$_3$ \cite{Manaka2} is regarded as quasi-1D S=1/2
FM-AFM alternating material, and
(CH$_3$)$_2$CHNH$_3$CuBr$_3$ \cite{Manaka3} is a S=1/2 AFM
dimerized material. Mixing these two compounds, they observed that
a gapless phase appeared in the regime of the intermediate concentration
$0.44<x<0.87$ of FM bonds by measuring magnetic susceptibility and
specific heats. In order to describe the properties of this
material, Hida \cite{Hida} and Nakamura
\cite{Nakamura} suggested a 1D model
\begin{equation}
H=\sum^N_{i=1}J{\bf S}_{2i-1}\cdot{\bf S}_{2i}+\sum^N_{i=1}J_i{\bf
  S}_{2i}\cdot{\bf S}_{2i+1},
\label{Hamiltonian}
\end{equation}
where ${\bf S}_i$ presents a spin S=1/2, $J>0$, $J_i=J_F(<0)$ with
a probability $p=x^2$  and $J_i=J_A(>0)$ with $1-p$.
This model has two limits: (i) when $p=1$, it is a S=1/2 AFM-FM
alternating Heisenberg spin chain. When $\mid J_i/J \mid > 1$ its
ground state is the Haldane phase with gapped energy spectrum;
(ii) when $p=0$, it is a dimerized AFM Heisenberg spin chain, its
ground state is the singlet dimer phase with gapped spectrum.
By density matrix renormalization group (DMRG) method, Hida
\cite{Hida} considered the case $p \geq 0.6$ with $J=1.0$ and $\mid J_i
\mid =2.0$ or $4.0$, and confirmed that there exists QG singularity when $p \leq
0.7$. Nakamura \cite{Nakamura} studied
the model by non-equilibrium relaxation analysis of the
quantum Monte Carlo (QMC) simulation in the whole parameter space
of concentration $0 \leq x \leq 1$ with $J=1.0$ and $\mid J_i \mid
=2.0$, and found the gap vanishes in the regime $0.44 <x<0.87$
consistent with the experimental results \cite{Manaka1}. However, in
all these numerical works, the finite temperature magnetic properties
in the whole parameter space of $x$ are absent. In order to directly compared
with the experimental results \cite{Manaka1}, we perform the finite
temperature QMC simulations extensively on this model in this paper.
% as a starting point to
%compare with the experiment measurements.

\section{QMC simulation results}
We investigate the magnetic and thermodynamic properties of the model
defined by eq. (\ref{Hamiltonian}) with $J=1.0$ and $\mid J_i \mid =
2.0$ using QMC simulations of continuous imaginary time loop/cluster algorithm.
The weak AFM coupling $J=1.0$ is fixed on all odd position bonds.
For even position bonds, the strong AFM or FM bonds $\mid
J_i \mid =2.0$ are chosen randomly according to the probability $p$. We perform
simulations for $100 \sim 200$  bond arrangement configurations. For
each bond configuration, after $2000$ Monte Carlo
steps (MCS) for thermalization,  we further update $2000$ MCS for
Monte Carlo average. The temporal and spatial periodic boundary
conditions are chosen for all simulations. In order to convince us of
the code validity, we study S=1/2 dimerized AFM chain in weak dimerization, 
and find our results for the energy gap consist well with
recent DMRG results \cite{Papenbrock}.

\begin{figure}[ht]
\begin{minipage}[t]{4.8cm}
\epsfxsize \epsfysize
\centerline{\psfig{file=affm-zd.eps,width=2in}} 
\vspace*{4pt}
\caption{The energy gap $\Delta$ and VBS order parameter $Z_L$
versus the probability $p$.}
\label{zd_eps}
\end{minipage}
\hspace*{70pt}
\begin{minipage}[t]{5cm}
\epsfxsize \epsfysize
\centerline{\psfig{file=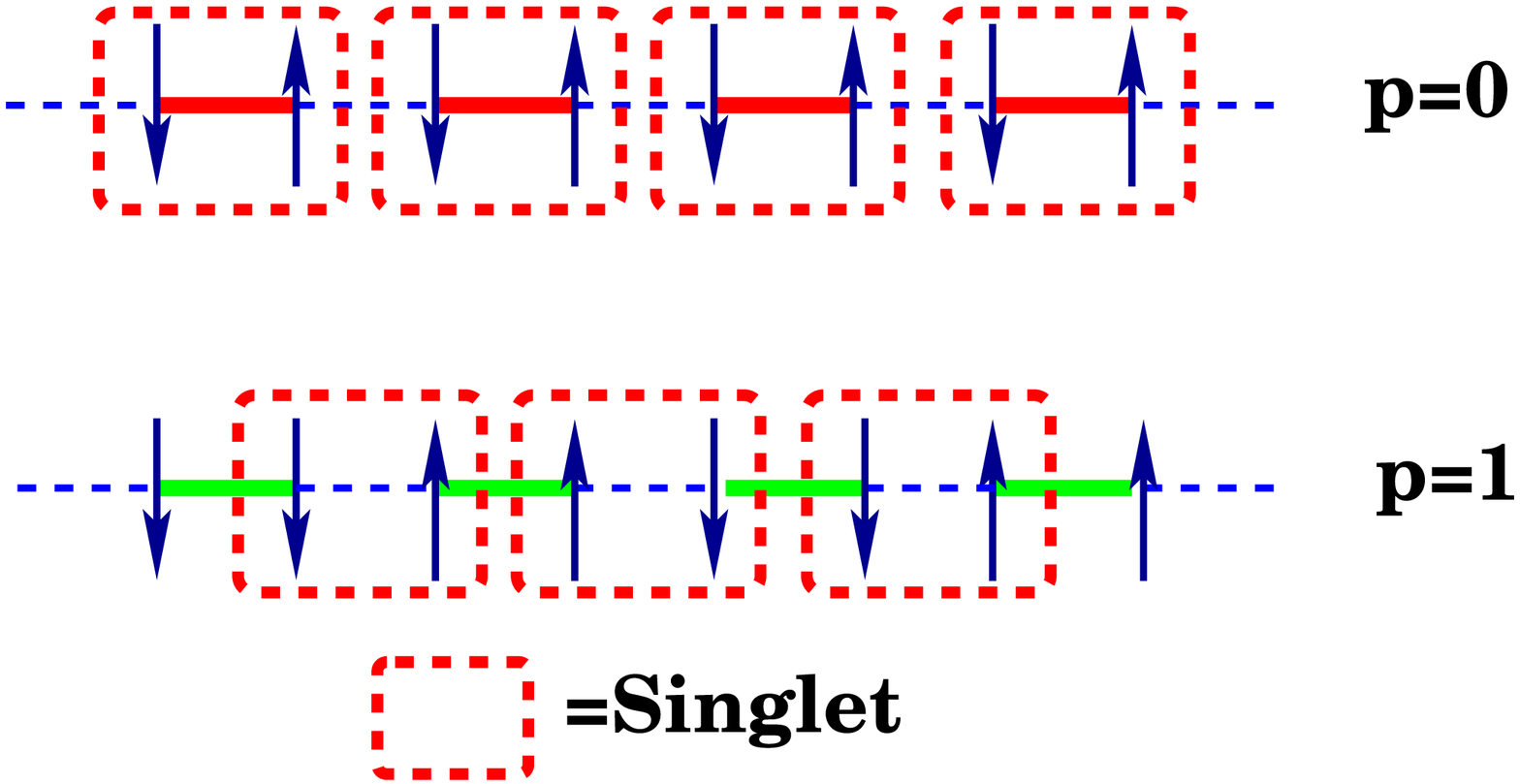, width=2.4in}}
\vspace*{4pt}
\caption{The illustration of spin ordered phases with the probability
  $p=0$ and $1.0$.}
\label{limitsfig}
\end{minipage}
\end{figure}

We first investigate the ground state properties on the system size
$L=128$ for temperatures as low as $\beta=1/T=200$. At low
temperature, the energy gap $\Delta$ is estimated by
\begin{equation}
\Delta = \lim_{L \rightarrow \infty}\frac{1}{\xi_{\tau}},
\end{equation}
where $\xi_{\tau}$ is the imaginary time correlation length
obtained by second-moment method \cite{Todo1}. Then the
valence-bond-solid (VBS) order parameter \cite{Todo2}
\begin{equation}
Z_{L} = <{\rm exp}[i\frac{2\pi}{L}\sum^{L}_{j=1}jS^{z}_{j}]>.
\end{equation}
is measured to distinguish different ground state phases. The
results obtained are shown in Fig. \ref{zd_eps}

It is interesting to find that energy gap $\Delta$ exists for the
Haldane phase at $p=1$ point. It gradually approaches to zero at $p
\approx 0.7$, and it opens again at very small $p \approx 0.02$.
It's hard to locate accurately the vanishing point of $\Delta$ because our $\Delta$ results are
not exact zero value due to finite-size effects. But our calculations
show that the gap should close at $p_{c1} \approx 0.02$ and $p_{c2}
\approx 0.7$. As a result, we presume that the system stays in the
gapless phase in the regime of $ p_{c1} < p < p_{c2}$, which is consistent with the
previous results \cite{Nakamura}.

In Fig. \ref{zd_eps}, we also observe that $Z_L \approx 1.0$ at
the limit $p=1$, keeps at finite in the regime $0.35 < p < 1.0$,
changes its sign at $p \approx 0.35$,  and then turns down to $-1.0$ when
$p <0.35$. The values of $Z_l \approx \pm 1.0$ characterize the two
limits of different ordered phases presented in Fig. \ref{limitsfig}.
%\begin{figure}[ht]
%\epsfxsize \epsfysize
%\centerline{\psfig{file=affm_limit.eps,width=2.0in}}
%\vspace*{}
%\caption{The illustration of spin ordered phases with the probability
%  $p=0$ and $1$.}
%\label{limitsfig}
%\end{figure}

Recent QMC study \cite{Arakawa} on S=1 random bond-alternating Haldane chain
 has shown that the VBS order parameter $Z_L$ is not effected by QG
 singularity, and it is an effective parameter to locate the RS
 critical point. Combining the results of $Z_L$ and $\Delta$, we
find that in the regime of $0.02 < p < 0.35$ where $\Delta$ vanishes and
$Z_L$ approaches zero from finite values, the system belongs to a
 critical phase. At the point $p \approx 0.35$, as $Z_L$ changes its
 sign, a phase transition happens. In the regime of $0.35<p<0.7$,
 $\Delta$ vanishes and $Z_l$ increases from zero to finite values.
 This fact reveals the system enters to other critical phase. In the
 regime of $0.7<p<1.0$, where both $\Delta$ and $Z_L$ are
finite, the system keeps in an ordered phase.

In order to distinguish the upper mentioned different phases, we
calculate the uniform magnetic susceptibility $\chi_u$ over the whole
parameter space $0<p<1.0$ at finite
temperatures. The results are summarized as following.

\begin{figure}[ht]
\begin{minipage}[t]{5cm}
\centerline{\psfig{file=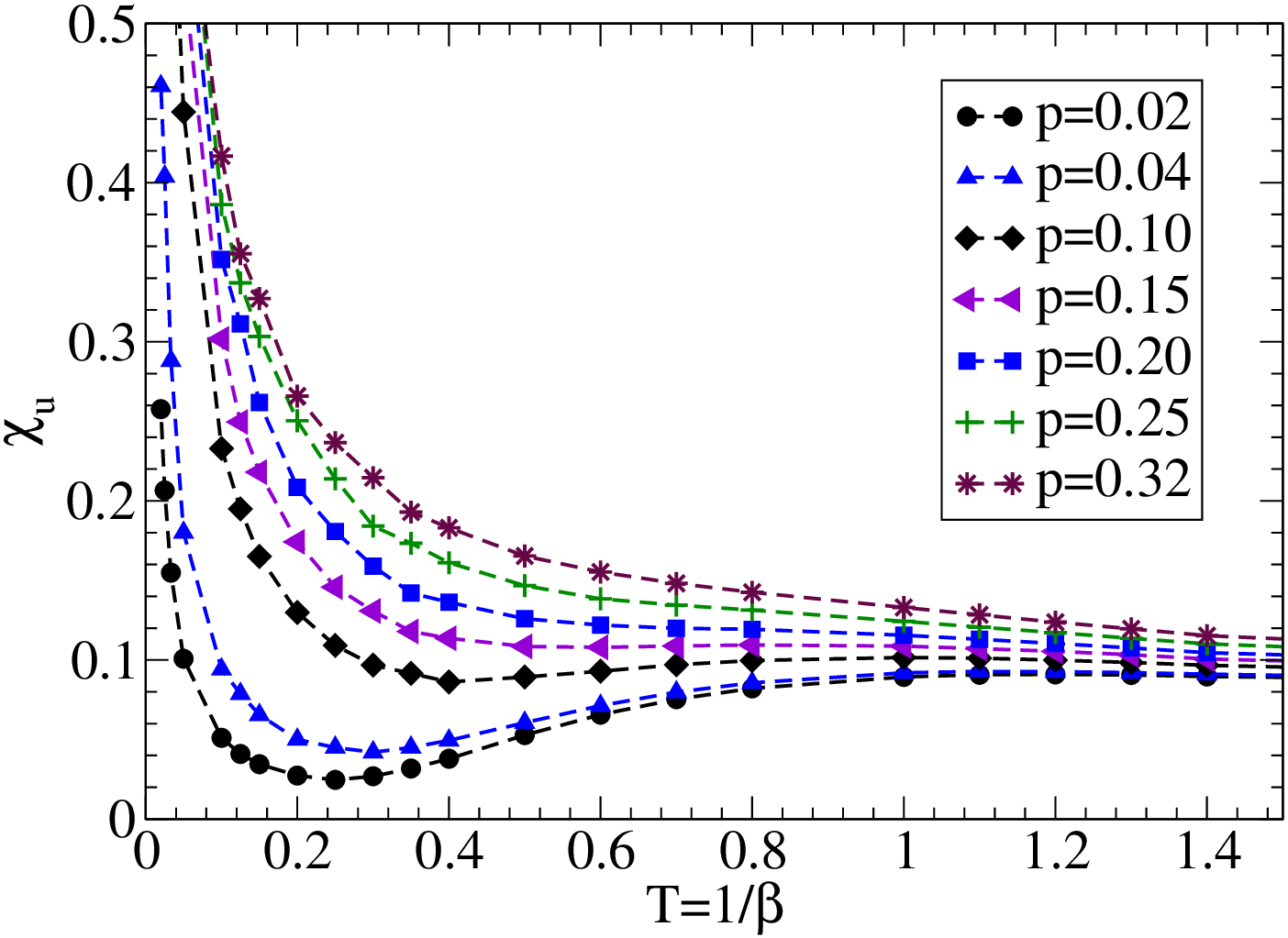,width=2.0in}} \caption{The
uniform susceptibility at $0.02<p<0.35$ versus temperature
  $T=1/\beta$.}
\vspace*{4pt}
\label{xu04_eps}
\end{minipage}
\hspace*{70pt}
\begin{minipage}[t]{5cm}
\centerline{\psfig{file=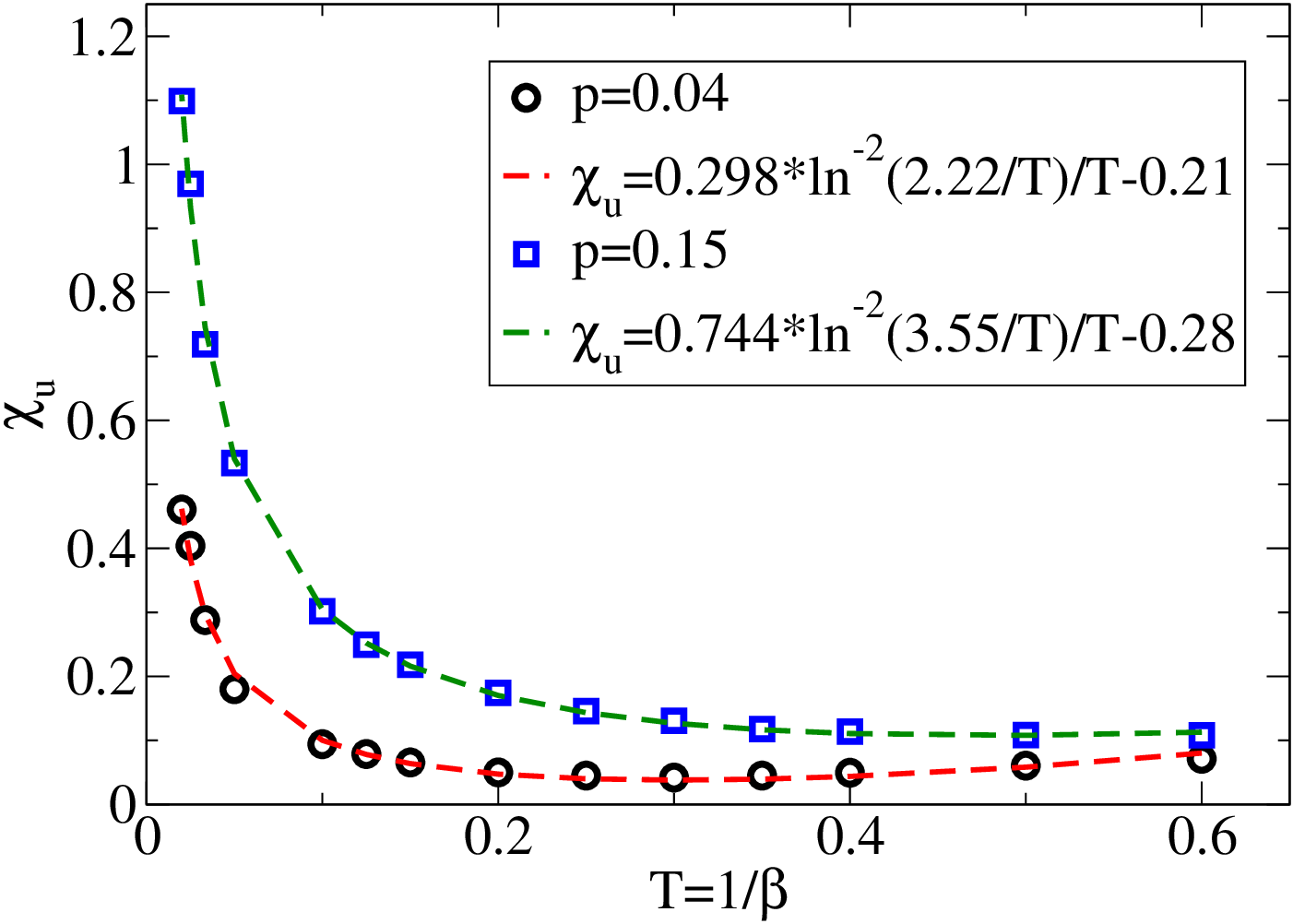,width=2.0in}}
\vspace*{4pt}
\caption{The fitness of $\chi_u$ by $\frac{{\bf ln}^{-2}(\Omega/T)}{T}$}
\label{fitRS_eps}
\end{minipage}
%\vspace*{4pt}

\end{figure}

{\it I.} As shown in Fig. \ref{xu04_eps}, 
in the regime $0.02<p<0.35$, the system keeps in gapless phase, where $\chi_u$
diverges when $T \rightarrow 0$, and every $\chi_u$ curve appears a
valley, that is the typical feature of eq. (\ref{eq_RSxu}) for
denoting RS phase. We fit the curves of $p=0.04$ and $p=0.15$ by eq.
(\ref{eq_RSxu}), and find they can be very well fitted as plotted in
Fig. \ref{fitRS_eps}. Thus we believe that the phase in this regime
belongs to RS phase.

\begin{figure}[ht]
\begin{minipage}[t]{5cm}
\centerline{\psfig{file=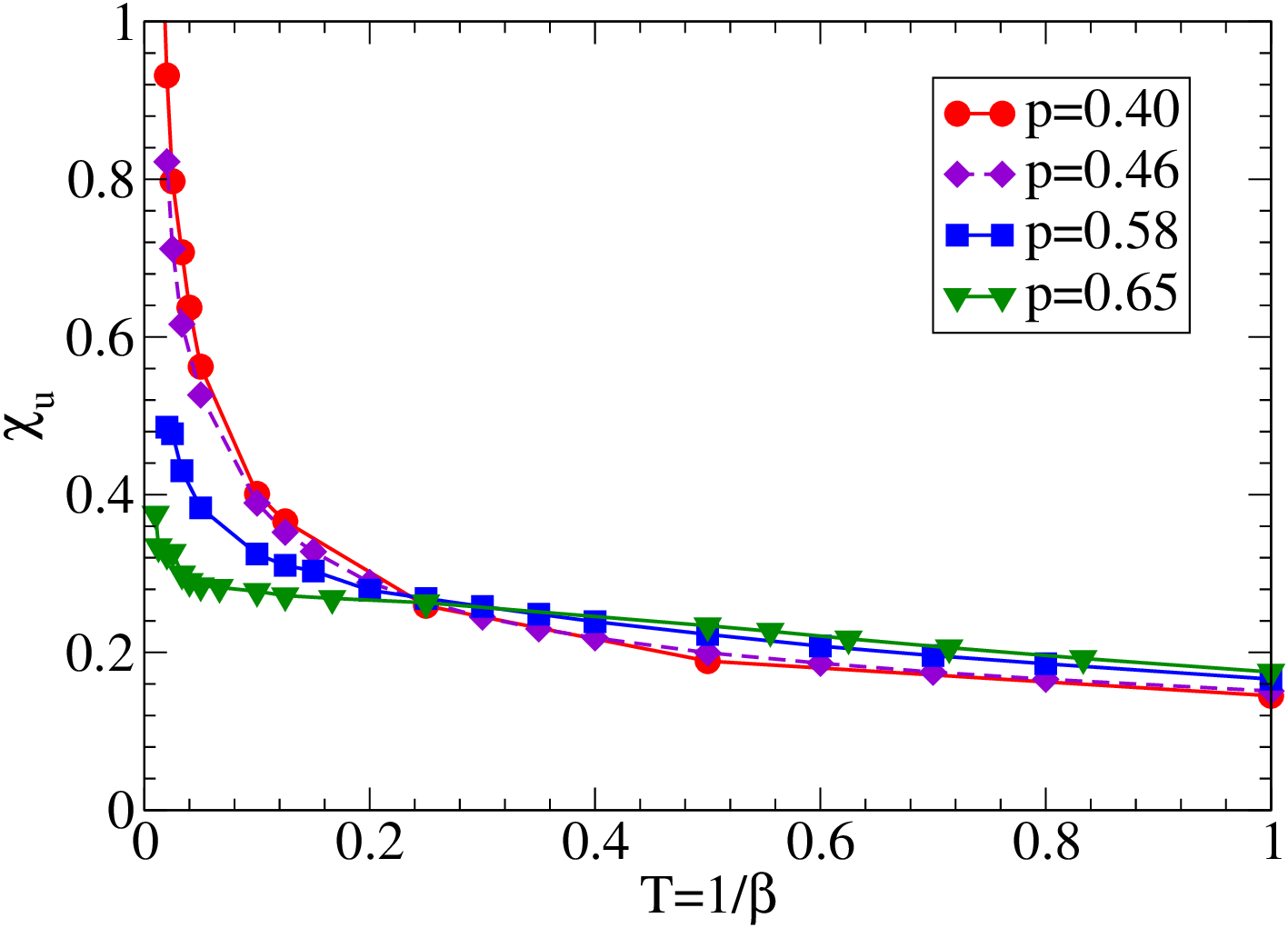,width=2.0in}}
\vspace*{4pt}
\caption{The uniform susceptibility at $0.35<p<0.7$ versus
temperature
  $T=1/\beta$.}
\label{xu0407_eps}
\end{minipage}\hspace*{70pt}
\begin{minipage}[t]{5cm}
\centerline{\psfig{file=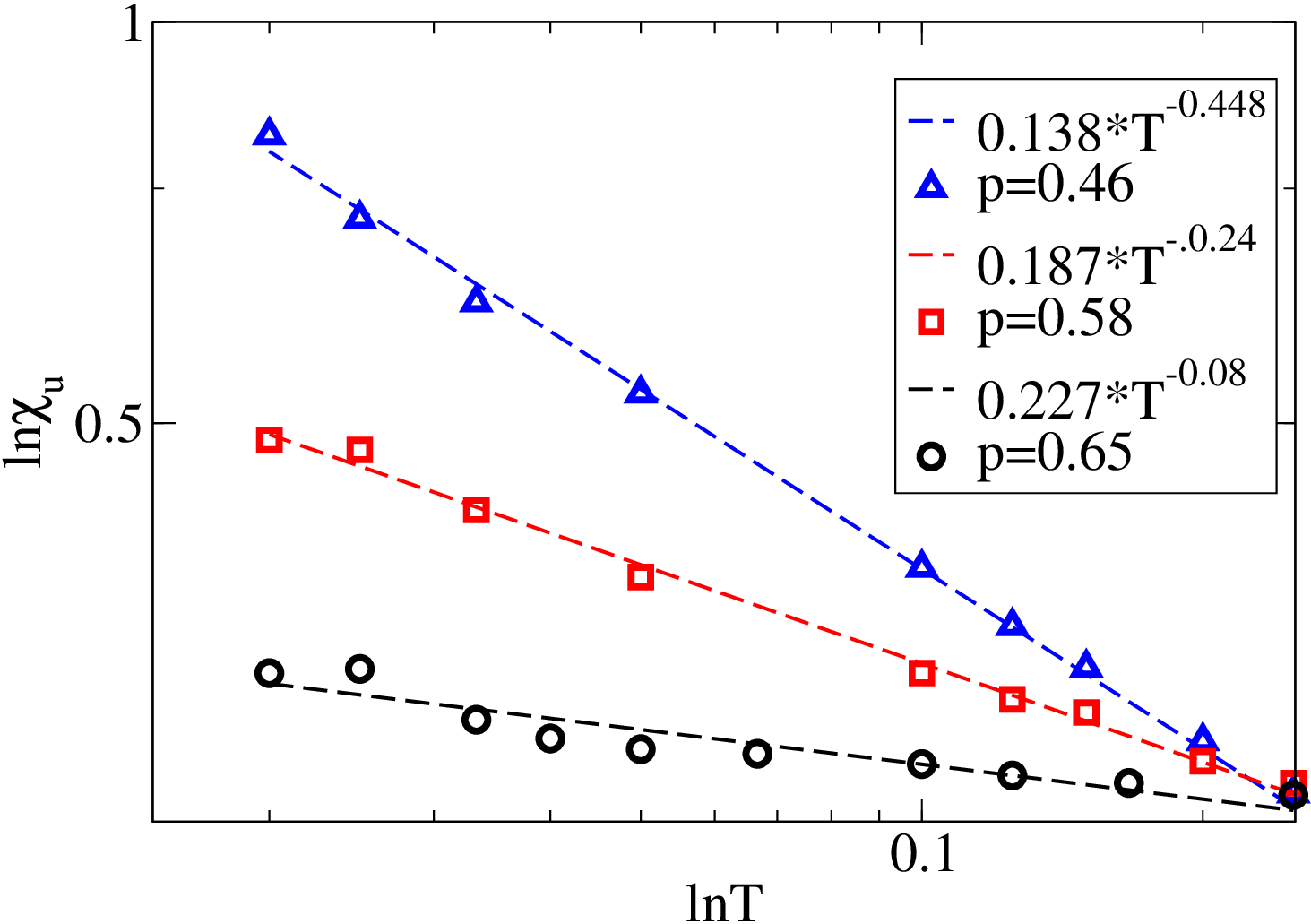,width=2.0in}}
\vspace*{4pt}
\caption{The fitness of $\chi_u$ by $T^{-\gamma}$}
\label{fitQG_eps}
\end{minipage}
\end{figure}

{\it II.} The regime of $0.35<p<0.7$ is also a gapless regime, where
$\chi_u$ curves diverge too, but they are obviously different from
those in the regime of $0.02<p<0.35$. Instead, as plotted in
Fig. \ref{xu0407_eps}, these curves are very similar as those in the
QG phase, where the typical feature is described by eq. (\ref{eq_QGxu}).
 In Fig.\ref{fitQG_eps}, we fit our low temperature results of
$\chi_u$ by $T^{-\gamma}$, and find again the fitness are
quite good. It is interesting to note that the behavior of $\chi_u$ is
not {\it Curie-like}, so one can believe the phase is not belong to the
universal class of AFM and FM bonds randomness \cite{Westerberg}. We
thus conjecture that the system is now in QG phase.

\begin{figure}[ht]
\begin{minipage}[t]{5cm}
\centerline{\psfig{file=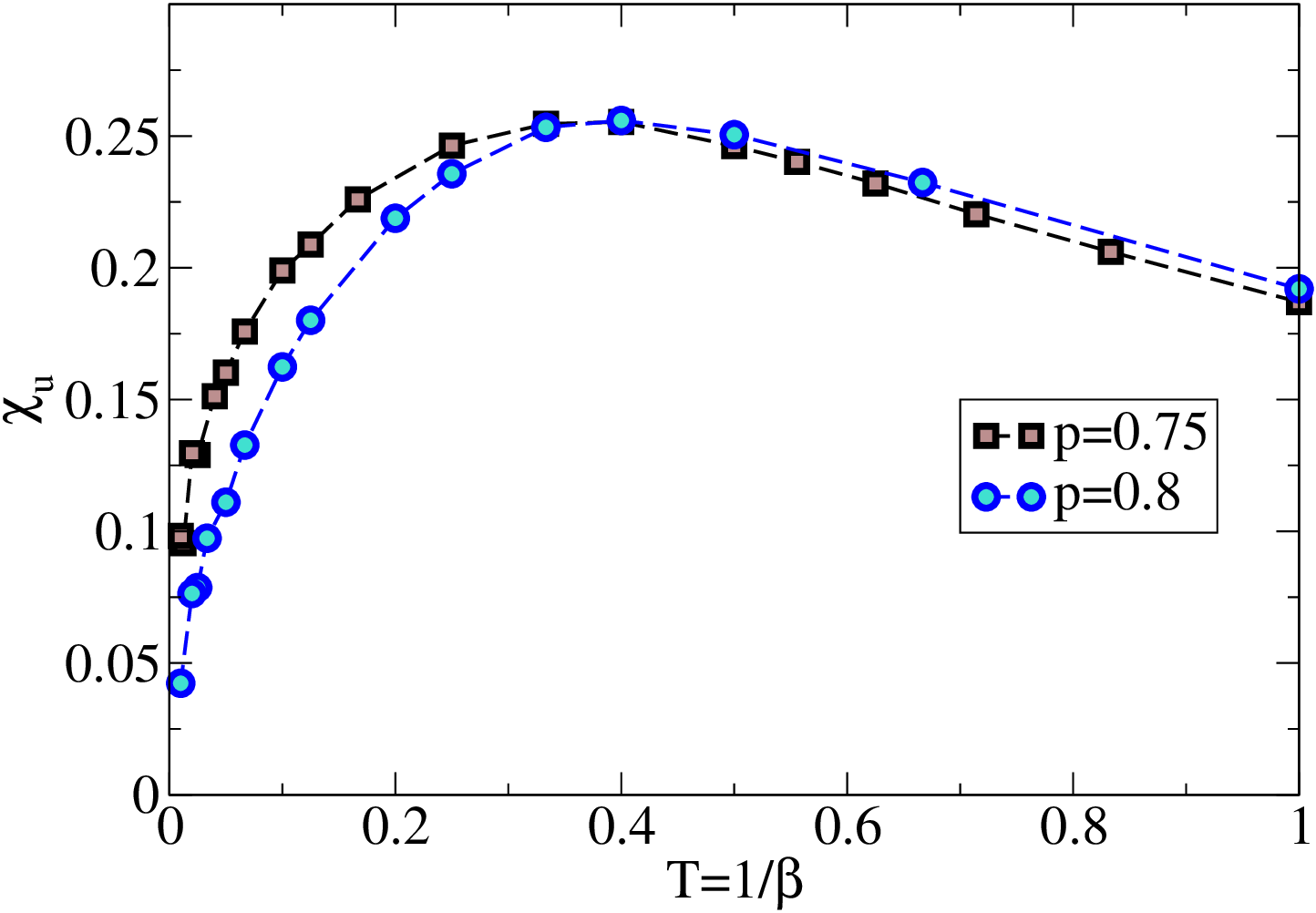,width=2.0in}}
\vspace*{4pt}
\caption{The uniform susceptibility at $0.7<p<1.0$ versus
temperature
  $T=1/\beta$.}
\label{xu0710_eps}
\end{minipage}
\hspace*{70pt}
\begin{minipage}[t]{5cm}
\centerline{\psfig{file=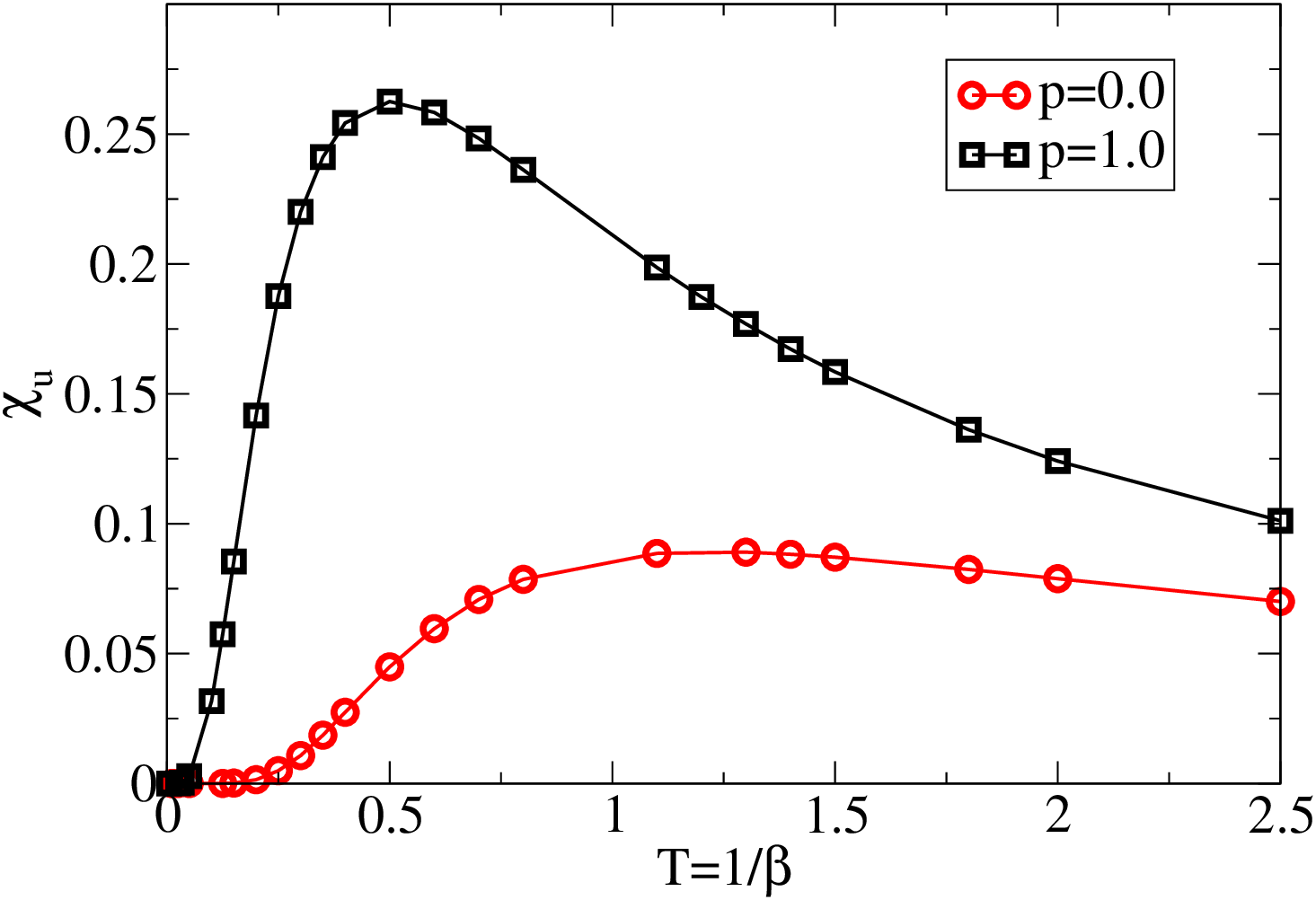,width=2.0in}}
\vspace*{4pt}
\caption{The uniform susceptibility at $p=0$ and $p=1.0$ versus
temperature
  $T=1/\beta$.}
\label{xulimit_eps}
\end{minipage}
\end{figure}

{\it III.} For the regime of $0.7<p<1.0$, the system enters to a gapped
phase because all $\chi_u$ curves appear the tendencies going to zero
when $T \rightarrow 0$. Our results are
plotted in Fig. \ref{xu0710_eps}.

{\it IV.} At last, we consider the two limit cases $p=0$ and $p=1$,
and our results are plotted in Fig. \ref{xulimit_eps}. Obviously, the
system belong to the gapped Haldane phase and
dimerized AMF phase on these two points, respectively.

\section{Conclusion and discussion}
From our QMC calculations, we can conclude that this S=1/2 AFM-FM
alternating bond randomness chain has four different phases with
respect to the probability $p$: ({\it i}) $p=0$, the
system is a dimerized AFM chains with gapped energy spectrum;
({\it ii}) in the regime of $0.02<p<0.35$, the system enters to the RS
phase, whose energy spectrum is gapless and the uniform magnetic
susceptibility $\chi_u$ obeys the eq. (\ref{eq_RSxu}); ({\it iii})
in the regime of $0.35<p<0.7$, the system turns to the QG phase where the
energy gap vanishes and the curves of $\chi_u$ consist with eq. (\ref{eq_QGxu});
({\it iv}) in the regime of $0.7<p \le 1.0$, the system is again in a
gapped phase. Finally, the case $p=1$ corresponds to the gapped AFM-FM alternating
spin chain.

Consequently, there should be three phase boundaries between these different phases:
({\it i})$\rightarrow$({\it ii}), because there is no effective
quantities to locate the exact position of this boundary, we 
only say that the transition from the dimerized phase to the RS phase
happens at very small $p \approx 0.02$; ({\it ii})$\rightarrow$({\it
  iii}), the phase boundary between RS phase and the QG phase resides at
$p=0.35$, where both the results of VBS order parameter $Z_L$ and
susceptibility $\chi_u$ consist reciprocally. $Z_L$ changing its sign
at the point $p=0.35$ also implies that it is a good quantity to indicate
the transition from RS phase to other phase, which confirm the
previous argument \cite{Arakawa}; ({\it iii})$\rightarrow$({\it
iv}), the location of the boundary between QG phase and the gapped phase at
$p \approx 0.7$ is hard to be determined by results of energy gap $\Delta$ because
of the finite-size-effects, but it can be extracted from the
behaviors of $\chi_u$ at different temperature. In the QG
phase, $\chi_u$ diverges as $T \rightarrow 0$, but in a gapped
phase, $\chi_u$ approach zero when $T \rightarrow 0$. Thus there
should be a cross point of $\chi_u$ at different temperature which
correspond to the boundary. We investigate the case of
same system size $L=128$ at different temperature $1/T=\beta
=10,50,100,200$, and analyze the results by finite-size-scaling of
imaginary time \cite{Sachdev}.
\begin{figure}[ht]
\centerline{\psfig{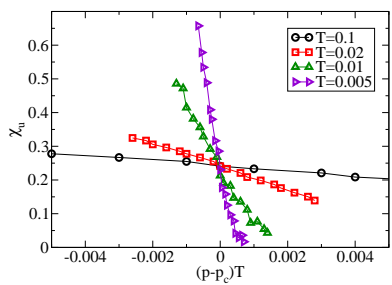}}
\vspace*{4pt}
\caption{Finite size scaling by $T=1/\beta$ of the uniform susceptibility near $p_c=0.71$.}
\label{scalex_eps}
\end{figure}
Our results, plotted in Fig. \ref{scalex_eps}, have shown that the
 gap should close at $p_c=0.71 \pm 0.01$, this result consist with
previous DMRG results \cite{Hida}.

This model can be described under the scenario for the study of
S=1 AF bond random chain by R. Hyman and K. Yang \cite{Hyman1}, and 
our result give a positive evidence to support their conclusion. More
important, our QMC results of $\chi_u$ are consistent with the experiment
results \cite{Manaka1} of quasi-1D (CH$_3$)$_2$CHNH$_3$Cu(Cl$_x$Br$_{1-x}$)$_3$
in the two limits $p=0,1$, as found in Fig.\ref{xulimit_eps}. In the
gapless regime, $0.02<p<0.71$, where $\chi_u$ diverge when $T
\rightarrow 0$, the behavior are similar with the experiments
observation that $\chi_u(T)$ diverges also for $0.56<x<0.83$ at low
temperature. Especially, we find the critical point $p_{c2}=0.71$ corresponding
to $x=0.84$ is very close to the experimental $x=0.83$. As the critical point
$p_{c1} \approx 0.02$ corresponding to $x \approx 0.14$ is different from the
experimental $x=0.56$, it implies the current single chain model
eq. (\ref{Hamiltonian}) is failed to describe the results of
experiment for small $p$, and some additional terms such as the weak
inter-chain coupling should be taken into account. Furthermore, we
believe that the behavior of $\chi_u(T)$ in the regime $0.56< x
<0.87$ for the experiment should successively exhibit first
the QG-type divergence $T^{-\gamma}$, then the RS-type divergence
${\bf ln}^{-2}(\Omega/T)/T$ at low temperatures. Our further
simulations are under consideration, and interesting results are expected in the
near future.

\section*{Acknowledgments}
Xu thanks for valuable discussion with Dr. H. Huang and Dr. P. Crompton.
 This work was supported in part by the NNSF of China and NSF of Zhejiang
province.


\begin{thebibliography}{0}
\bibitem{Dasgupta} C. Dasgupta and S.-K. Ma, Phys. Rev. B {\bf 22} 1305(1980).

\bibitem{Fisher} D.S. Fisher, Phys. Rev. B, {\bf 50},3799(1994).

\bibitem{Hyman1} R.A. Hyman and K. Yang, Phys. Rev. Lett. {\bf 78},
  1783(1997).

\bibitem{Hyman2} R.A. Hyman, K. Yang, R.N. Bhatt and S.M. Girvin,
  Phys. Rev. Lett. {\bf 76}, 839(1996).

\bibitem{Uchiyama} Y. Uchiyama, Y. Sasago, I.Tsukada, K. Uchinokura,
  A. Zheludev, T. Hayashi, N. Miura, and P. Boni, Phys. Rev. Lett.
  {\bf 84}, 632(1999).

\bibitem{Azuma} M. Azuma, Y. Fujishiro, M. Takano, M. Nohara and
  H. Takagi, Phys. Rev. B {\bf 55}, R8658(1997).

\bibitem{Masuda} T. Masuda, A. Zheludev, K. Uchinokura, J.-H. Chung
  and S. Park, cond-matt/0404688.

\bibitem{Damle} O. Motrunich, K. Damle and D.A. Huse, Phys. Rev. B
{\bf 63}, 134424(2001).

\bibitem{Westerberg} E. Westerberg, A. Furusaki, M. Sigrist and
  P.A. Lee, Phys. Rev. Lett. {\bf 75}, 4302(1997).

\bibitem{Frischmuth} B. Frischmuth, M. Sigrist, B. Ammon and
  M. Troyer, cond-matt/9808027.

\bibitem{Nguyen} T.N. Nguyen, P.A. Lee, and H.-C. zur Loye, Science
  {\bf 271}, 489(1996).

\bibitem{Manaka1} H. Manaka, I. Yamada and H.A. Katori, Phys. Rev. B
  {\bf 63}, 104408(2001).

\bibitem{Manaka2} H. Manaka, I. Yamada and K. Yamaguchi,
  J. Phys. Soc. Jpn. {\bf 66}, 564(1997).

\bibitem{Manaka3} H. Manaka, I. Yamada, J. Phys. Soc. Jpn. {\bf 66},
  1908(1997).

\bibitem{Hida} H. Hida, J. Phys. Soc. Jpn. {\bf 72}, 688(2003).

\bibitem{Nakamura} T. Nakamura, J. Phys. Soc. Jpn. {\bf 72},
  789(2003).

\bibitem{Todo1} S. Todo and K. Kato, Phys. Rev. Lett. {\bf 87}, 047203(2001).

\bibitem{Todo2} M. Nakamura and S. Todo, Phys. Rev. Lett. {\bf 88},
  167208(2002).

\bibitem{Papenbrock} T. Papenbrock, T. Barnes, D.J. Dean,
  M.V. Stoitsov, and M.R. Strayer, Phys. Rev. B {\bf 68},
  024416(2003).

\bibitem{Arakawa} T. Arakawa, S. Todo and H. Takayama,
  cond-mat/0410755.

\bibitem{Sachdev} S. Sachdev, {\it Quantum Phase Transition}(1999),
  Cambridge University Press.

\end{thebibliography}
\end{document}